\makeatletter \renewcommand\@biblabel[1]{{#1}. } \makeatother
\documentstyle[epsfig,12pt,doublespace,overcite]{article} 
\begin{document}
\title{\textbf{Sidechain Dynamics and Protein Folding}}
\author{Edo Kussell$^{1}$, Jun Shimada$^{2}$, Eugene I. Shakhnovich$^{2\star}$}
\date{\today}
\maketitle
\begin{center}
\vspace{0.25in}
\large
\textbf{Running title:} Sidechain Dynamics\\
\normalsize
\vspace{0.55in}
48 pages (including figure/table captions), 13 figures, 3 tables.\\
\textbf{manuscript:} ``sc dynamics.tex'' (LaTeX document) \\
\vspace{0.5in}
\begin{spacing}{1}
$^{1}$Department of Biophysics \\
Harvard University \\
240 Longwood Ave. \\
Boston, MA 02115 \\\indent \\
$^{2}$Department of Chemistry and Chemical Biology \\
Harvard University \\
12 Oxford Street  \\
Cambridge MA 02138\\
\vspace{0.5in}
$^\star$corresponding author \\
tel: 617-495-4130 \\
fax: 617-496-5948 \\
email: eugene@belok.harvard.edu \\
\vspace{0.5in}

\end{spacing}
\end{center}

\newpage
\noindent
\textbf{Abstract} \\ The processes by which protein sidechains reach
equilibrium during a folding reaction are investigated using both
lattice and all-atom simulations.  We find that rates of sidechain
relaxation exhibit a distribution over the protein structure, with the
fastest relaxing sidechains being involved in kinetically important
positions.  Traversal of the major folding transition state
corresponds to the freezing of a small number of residues, while the
rest of the chain proceeds towards equilibrium via backbone
fluctuations around the native fold.  The post-nucleation processes by
which sidechains relax are characterized by very slow dynamics, and
many barrier crossings, and thus resemble the behavior of a glass.  At
optimal temperature, however, the nucleated ensemble is energetically
very close to equilibrium; slow relaxation is still observed.  At
lower temperatures, sidechain relaxation becomes a significant and
very noticeable part of the folding reaction.
\noindent

\vspace{1in}

\newpage
\section*{Introduction}

Protein folding is a complex, single molecule process in which a
polypeptide backbone with diverse sidechain groups efficiently
searches a vast conformational space and finds a unique native fold.
Most theoretical attempts to understand the folding process have
modeled the polymer, in one way or another, as a chain of beads which
undergoes a backbone freezing transition.  The internal degrees of
freedom of each sidechain (the $\chi$ angles) add another layer of
difficulty to understanding the folding process.  

In unfolded conformations, the barriers between rotamer states of
sidechains are low~\cite{kn:proteins_brooks} and sidechains easily
convert between them.  Upon folding, buried sidechains are usually
found in a single, well-defined rotamer
state~\cite{kn:richards_lim,kn:non_rotamers,kn:levitt_endgame,kn:dunbrack_karplus},
and interconversion between rotamers, when energetically allowed, is
slow due to high barriers~\cite{kn:fersht,kn:shakh_finkel}.  Because
protein folding is thought to be a sidechain-driven process, finding
the native rotamers is an integral part of the folding reaction.  Do
sidechains reach their native conformations simultaneously with the
backbone, or is sidechain ordering a separate process that occurs
after the native fold has been reached?  This question poses a
challenge for experimentalists and theoreticians alike.

We study the dynamics of sidechains during the folding process via a
simplified model, which captures the basic physical aspects of the
problem, as well as an all-atom protein folding simulation, in which
the full complexity of sidechain shapes and mobility is represented.
We start by modifying the classic lattice model of proteins to account
for internal sidechain states of each amino-acid.  This new model,
like its predecessors, is found to fold in a cooperative all-or-none
manner.  The presence of internal degrees of freedom at each site,
however, allows the backbone to reach the native conformation before
all of the sidechains have become properly ordered.  Since energy
comes from sidechain-sidechain interactions, completion of the folding
reaction requires both the backbone and the sidechains to reach their
native states.  We find that, depending on simulation temperature, the
backbone can become native-like long before the internal sidechain
states reach equilibrium.  By measuring the rate of sidechain ordering
for each monomer, we find that there is a cluster of residues whose
sidechains become ordered very fast, while the relaxation rates of
other positions is up to an order of magnitude slower.  The fast
cluster turns out to be very close to the folding nucleus identified
previously for the structure we use.

A small number of kinetically important residues thus freeze in their
native ``rotamer'' state on a fast timescale, pulling the backbone
strongly toward its native conformation, while the rest of the protein
relaxes on a slower timescale toward its equilibrium energy.  This
slow phase could not be observed in previous lattice simulations
because it arises entirely from the presence of the internal sidechain
states of each monomer.  We find that the slow-phase relaxation of
energy to equilibrium follows stretched-exponential kinetics,
suggesting that the dynamics are exhibiting some glass-like properties
due to sidechains.  We note that backbone-only lattice models have
been shown to be free of a glass transition over a wide range of
temperatures~\cite{kn:lattice_glass}.  Various experimental and
theoretical studies have suggested that some aspects of protein
folding might be interpreted as glassy
behavior~\cite{kn:wolynes_glass,kn:brooks_glass,kn:karplus_glass,kn:wand_glass},
though whether these phenomena are attributable to backbone,
sidechains, or solvent remains to be seen.

Because the lattice model is computationally very fast, one can
observe relaxation to equilibrium even at lower temperatures.  This
becomes impossible once more realistic models are used.  Our previous
work using an all-atom simulation, with a simplified G\={o}
potential~\cite{kn:go_potential}, demonstrated that folding to the
native backbone topology (rms deviation $<$ 1 \AA\ from crystal
structure) happened within a reasonable amount of time, but that full
relaxation to equilibrium could not be observed at temperatures below
a certain threshold.  Thus the slow phase observed in the lattice
model is also present in the all-atom model, but cannot be fully
characterized therein due to prohibitively long simulation times.
This does not prevent us from identifying the residues that exhibit
fast transition to the native state, because the fast phase in which
the native backbone conformation is reached is fully accessible to our
simulation.  We characterize the folding transition state of Protein G
using the all-atom simulation.  We also identify the residues whose
sidechains exhibit fast relaxation to their native state.  We find
that these same residues play a key role in the transition state
ensemble.  As in the simpler lattice model, we find a wide
distribution of sidechain relaxation times.  Two very different models
are thus in marked agreement; together they provide a clear picture of
sidechain dynamics during the folding process.

\section*{Results}
In order to mimic the internal degrees of freedom of protein
sidechains (the $\chi$ angles), we modified the standard lattice model
by adding $n$ sidechain states to each residue.  This is consistent
with the observation that protein sidechains usually populate discrete
rotameric configurations~\cite{kn:richards_lim,kn:non_rotamers}.  The
state of each residue at any given time is a number between 0 and
$n-1$. Of these $n$ states only one state (the 0 state) was designated
as native for each residue.  When two residues came into contact
during simulation, they interacted only if both were in their native
state - a contact formed with one or both residues in a non-native
sidechain state did not contribute to the energy of the conformation
(see Methods).  While there are other ways to model a native
vs. non-native rotamer interaction using a lattice model (for example,
we could have assigned some fraction of the native energy when
non-native monomers interact), we chose the present scheme for
simplicity.  Previous lattice models have added sidechain degrees of
freedom by letting sidechains occupy a lattice
site~\cite{kn:sidechains_dill,kn:sidechains_thirumalai,kn:sidechains_lewyn}.
In our model, sidechain states are treated implicitly, resulting in a
considerable computational advantage.

An important aspect of sidechain motion in real proteins is that upon
compactification of the polypeptide chain, sidechain motion is
restricted due to the excluded volume
effect~\cite{kn:proteins_brooks,kn:packing_paper}.  In order for
sidechains to repack in the protein interior, the backbone must
perform a ``breathing motion''~\cite{kn:fersht}, allowing sidechains
some extra room to move, and thus making certain sidechain
configurations momentarily available.  Any model of sidechain dynamics
must incorporate this effect in some way.  Our all-atom simulation
contains this effect explicitly.  In the lattice simulation we mimic
this effect by our choice of moves.  In addition to the usual lattice
backbone moves, we allow the sidechain states of a given residue to
interconvert when there are no other residues in contact with it (see
Methods).  Thus, when the chain is fully compact, the sidechain states
are frozen until a backbone fluctuation frees some residues, and
allows their states to change.

We tested the lattice model with 1, 2, 4, and 8 internal states per
monomer using a 27-mer sequence designed to fold into a 3 x 3 x 3
cube.  The $n = 1$ model corresponds to the standard lattice model and
is shown here only as a control.  The thermodynamics of these four
models is shown in Figure \ref{fig:thermo}A.  All are seen to exhibit
a cooperative temperature transition, with the transition temperature
getting progressively lower as the number of internal states of each
monomer increases.  The lowered transition temperature is to be
expected as the increased entropy of the model (due to more internal
states) necessarily leads to some destabilization.  The transition
region becomes narrower as $n$ increases, due to the increase in
entropy of the unfolded state relative to the folded state.

We studied the kinetics of the various models by plotting the average
folding time as a function of temperature (see Figure
\ref{fig:kinetics}, panels A, B, and C).  The models with $n > 1$
possess the property that the backbone can reach full nativity before
all of the sidechains have become native.  This leads to the
interesting question of how the polymer chain reaches its native
energy.  That is, does the formation of the native backbone lead to
immediate sidechain ordering, or do sidechains relax slowly to
equilibrium after the chain has folded?

To answer this question, we plotted both the average time to reach the
native energy (which corresponds to full sidechain ordering), and the
average time to reach the native backbone in Figure
\ref{fig:kinetics}.  We find that at temperatures above the optimal
folding temperature, the native energy is reached immediately after
the native backbone is found, and thus sidechain ordering is fast.  At
low temperatures, on the other hand, the native backbone is reached
long before native energy is achieved, and sidechain ordering is slow.
The shift between these two behaviors appears to be continuous in
temperature, but depends on the number of sidechain states, $n$, per
residue.  For the $n = 2$ model (panel A), the shift from fast to slow
sidechain ordering occurs at the relatively low temperature of $12.5$
which is about 60\% of the temperature of fastest folding, $T_{opt}$.
For the $n = 4$ model (panel B), we see that the delay between
backbone folding and full ordering is noticeable even at $T_{opt}$,
and becomes significant at $T = 13.3$ which is 85\% of $T_{opt}$.  The
same is observed for the $n = 8$ model (panel C).  Thus the increase
in sidechain entropy of the chain leads to a severe sidechain-ordering
trap.  This trap becomes increasingly prominent as the entropy of the
model increases, and the temperature at which it becomes noticeable
moves closer and closer to $T_{opt}$.

The mechanism for reaching full nativity (backbone and sidechains) at
temperatures lower than $T_{opt}$ thus seems to be one in which the
native backbone structure is formed, followed by sidechain ordering
via backbone fluctuations around the native structure.  It is entirely
possible, however, that the native backbone structure is reached
during the folding trajectory but unfolds immediately because too few
sidechains are native.  This, in fact, is the case even at low
temperatures.  At some point in time, however, the native backbone
structure is reached with enough native sidechains to remain stable
long enough to allow the rest of the sidechains to become ordered.  It
is the ordering of sidechains after this \emph{stable} native backbone
is reached that we identify as an important kinetic step at
temperatures below $T_{opt}$.  Accordingly, we plot the average time
of the \emph{last} pass to the native backbone conformation in all of
our figures.  The time of the last pass is defined as the first time
the chain reached the native backbone without losing more than 50\% of
its native contacts before reaching the native energy.  We found that
our results did not change significantly when we varied the fraction
of native contacts used in this definition.

It is instructive to obtain a kinetic picture for an ``unhindered''
model in which the sidechain states of each residue can interconvert
freely, regardless of its surroundings.  Such a model represents
protein folding in molten globule conditions, in which sidechain
rotamers can easily interconvert.  We see in Figure
\ref{fig:kinetics}D that in the unhindered model, the slow ordering of
sidechains is not observed at any temperature.  Instead, once the
backbone reaches nativity, any non-native sidechains can immediately
become native, and they do because it is energetically favorable.
This control demonstrates, then, that the sidechain ordering trap is a
feature of folding under conditions in which a tight native state,
with low sidechain mobility, is the free-energy minimum.

We compare the kinetics of the unhindered model with $n = 2$ with the
kinetics of the standard lattice model ($n = 1$).  We find that at
their respective optimal folding temperatures, the average time to
reach the native energy for both models is the same (Figure
\ref{fig:kinetics}D).  The freely interconverting internal states,
then, do not have any effect on the kinetics of folding; they only
affect the stability of the model.  On the other hand, the folding
time of the hindered models with $n > 1$ at $T_{opt}$ is significantly
slower than that of the 1-state model.  This makes sense in light of
the observation that the hindered models are more difficult to
nucleate, for the following reason: if a proper nucleus forms and one
or more of its sidechains are in a non-native state, the nucleus will
almost certainly break apart because the energy of each of its
contacts is crucial for its ability to function as a nucleus
\cite{kn:specific_nucleus}.  The sidechains cannot interconvert while
the nucleus contacts are present, and therefore the nucleus dissolves.
In the unhindered model, the non-native sidechain states of a nucleus
can become native with the nucleus intact, so the folding time in this
model is not affected by the presence of sidechain states.

Having observed that folding to the correct backbone structure occurs
significantly before the native energy is reached, we asked the
following questions: Do some residues reach their native sidechain
state faster than others? If so which ones are fast, which ones are
slow, and why?  We decided to study a 48-mer structure whose folding
in the standard lattice model has been studied
exhaustively~\cite{kn:eugene_evolution}.  In order to maximize the
temperature range in which we could study folding of this structure,
we used a sequence that has been optimized for fast
folding~\cite{kn:lattice_evolution}.  In addition to having a fully
characterized nucleus, using a 48-mer sequence allowed us to see
whether our results were sensitive to the size of the structure.

The thermodynamics of the 48-mer sequence are shown in Figure
\ref{fig:thermo}B.  The kinetics for the 4-state model are shown in
Figure \ref{fig:48mer_kin}.  We consider a kinetic model with three
steps: \\ \indent \indent \indent Unfolded Backbone $\longrightarrow$
Folded Backbone $\longrightarrow$ Sidechain Ordering \\ Figure
\ref{fig:48mer_kin}B shows the average time of the first and second
steps plotted as diamonds and crosses, respectively.  We immediately
see that for high temperatures, the sidechain ordering step is several
orders of magnitude faster than the backbone folding step.  As
temperature becomes lower, the sidechain ordering time becomes
comparable to the backbone folding time.  At T = 0.13 = 85\%
$T_{opt}$, the rate of sidechain ordering becomes significant as it
comes within an order of magnitude of the rate of backbone folding.
The 48-mer sequence in the 4-state model, then, is seen to behave very
much like the 27-mer 4-state model, with both models developing a
significant sidechain ordering step in kinetics at 85\% of $T_{opt}$.

In order to obtain individual sidechain ordering rates for each
residue, we performed many long folding runs.  For each residue we
averaged its sidechain state over all folding runs: we assigned a
value of 1 to the native internal state of a given residue, and a
value of 0 to all other internal states, and at each timestep averaged
these values over runs.  Two traces obtained after averaging are shown
in Figure~\ref{fig:48mer_traces}.  We fit a single exponential (see
Methods) to each trace, and obtained time constants for each of the 48
residues.

The distribution of rate constants for two temperatures is given in
Figure \ref{fig:rate_hists}, and the fast residues are labelled by
number.  The first striking feature is that these distributions span
two orders of magnitude.  At the lower of the two temperatures (T =
7.4 = 81\% $T_{opt}$), most residues exhibit slow relaxation rates, as
seen by the sharp peak near zero.  At the higher temperature of
$T_{opt} = 9.1$, the height of the peak is reduced and more residues
are seen with faster rates.

At both temperatures, a small number of residues have very fast rates.
Many of these fast residues belong to the folding nucleus for this
structure that was identified in another
study~\cite{kn:eugene_evolution,kn:lattice_evolution} using the
standard lattice model.  In Figure \ref{fig:48mer_struct} we show the
48-mer structure colored by rate of sidechain freezing at $T = 7.4$,
and we indicate the original nucleus by large spheres.  Of the 10
fastest residues that become fully ordered at T = 7.4, 7 belong to the
folding nucleus.  While some of the nucleus positions are no longer
kinetically important in the present model, a strong signature of the
old nucleus has remained.  Importantly, with the exception of residue
9, all of the fast positions that reach full nativity are located in
or near the original nucleus was found.  

It appears, then, that at temperatures at or below $T_{opt}$, a small
group of residues reaches full nativity quickly, thus organizing a
critical piece of structure which remains fully stable, allowing the
rest of the chain to gradually order its sidechains.  At $T = 7.4$,
the formation of the stable piece traps many sidechains in non-native
states which take a very long time to reorganize via backbone
fluctuations.  On the other hand, at $T_{opt}$, as seen in Figure
\ref{fig:rate_hists}, more residues are found in the fast tail of the
rate distribution, indicating that backbone fluctuations are
sufficient to allow sidechain ordering to occur more quickly once
enough native structure has formed.  Additionally, the higher
temperature requires a larger amount of native structure to be formed
in order to remain stable.  These two effects act to eliminate
sidechain ordering as a relevant kinetic step at $T_{opt}$.  This is
seen clearly in Figure \ref{fig:48mer_kin} where at $T_{opt}$ the
sidechain ordering step is an order of magnitude faster than the
backbone folding step.

Another way to see that sidechain dynamics becomes markedly different
as temperature is lowered is given in Figure \ref{fig:energy_relax}.
The red line indicates the equilibrium energy at each of the two
temperatures, while the solid green line is a time trace of the
average energy over all runs.  The average time to form the stable
native backbone is 2.7 x $10^{7}$ at $T = T_{opt} = 9.1$, and 1.6 x
$10^{8}$ at $T = 7.4$, and is marked by an arrow in the figure.  For T
= $T_{opt}$, the arrow indicates that at the time of native backbone
formation, the energy of the chain is already very close to its
equilibrium value.  That is not the case at low temperature, at which
there is a significant gap between the energy of the folded chain and
the equilibrium energy.

We tried to fit the relaxation of energy by a standard, single-barrier
process (single exponential) as well as a double exponential fit -
both fits converged to the same curve which is shown as a dashed line
in Figure~\ref{fig:energy_relax}.  The fit is not appropriate at any
timescale.  In particular, we note that at short times, the trajectory
may resemble a single-exponential process, but it develops a very long
tail at long times.  We fit the long tail using a stretched
exponential, $b_1 \exp (-b_2 t^{\alpha})$, and found $\alpha = 0.09$;
the relaxation is therefore practically logarithmic at long times (see
Figure~\ref{fig:energy_relax} caption for details).  At short times we
fit a single-exponential.  These fits were done at $T = 9.1$ and are
shown as the solid black line in Figure \ref{fig:energy_relax}.  At $T
= 7.4$ we did not have enough data at long times to see relaxation to
equilibrium and therefore a fit would not be meaningful.  The fit of
energy relaxation using a single-exponential at short times, and a
stretched exponential at long times is very good.  The
single-exponential phase corresponds to the classic nucleation
mechanism in which the backbone topology becomes
organized~\cite{kn:specific_nucleus}.  The stretched exponential phase
is a signature of glassy dynamics associated with the sidechain
degrees of freedom and will be discussed below.

Because lattice models can give only a schematic view of the folding
process, we proceeded to investigate sidechain dynamics in an all-atom
simulation of Protein G, an alpha/beta protein that has featured in
numerous experiments~\cite{kn:igd_baker,kn:igd_redesign,kn:igd_roder}.
The details of the simulation and a full characterization of the
folding kinetics of this protein will be given elsewhere.  Our goal in
the present study is to see how the results obtained from our
simplified lattice model compare with a much more realistic
representation of a protein, and whether the same kind of analysis can
shed light on the kinetics of a real protein.  In the lattice model we
had to postulate a set of microscopic dynamics for the internal states
of each residue.  In the all-atom simulation, we model all sidechain
atoms and torsions explicitly.  We use the simulation methodology
described previously~\cite{kn:crambin_folding}.  Because rotations
around sidechain $\chi$ angles are continuous, interconversion between
sidechain rotamers can become restricted if a residue is buried.
Slowing down of sidechain dynamics upon compactification emerges from
the excluded volume interaction in this model, and does not have to be
included phenomenologically as in the lattice model.

We obtained 50 folding trajectories of Protein G, starting from random
backbone and sidechain conformations, all at the same temperature.
All runs were terminated after $2 \times 10^9$ steps, by which time 47
had reached the native backbone fold (see Methods).  We then applied a
time series analysis similar to the one we used for the lattice model.
Specifically, at each time step and for each residue we recorded a
value of 1 if the sidechain was in its native rotameric state, and a 0
otherwise.  We averaged these values for each residue over all
trajectories, and then fit a kinetic model to the resulting traces.
The parameters for the fits for each residue are given in Tables
\ref{tb:igd_fits2} and \ref{tb:igd_fits3}.

Protein G folds in simulations via a kinetic intermediate consisting
of either hairpin 1 and the helix or hairpin 2 and the helix (see
Discussion).  A two-state fit was therefore not appropriate for some
of the residues.  We used a three-state fit for all residues, and
found that for some of the residues one rate constant was at least an
order of magnitude larger than the other.  Such residues were
classified as two-state, while the others were classified as
three-state.  Figure \ref{fig:igd_traces} shows representative fits
for two-state and three-state residues.  The relaxation rates given in
Tables \ref{tb:igd_fits2} and \ref{tb:igd_fits3} span an order of
magnitude.  The equilibrium level of ordering of each residue
(parameter $d$ in the Tables) was obtained directly as an average over
a long simulation started in the native state, and was not obtained by
fitting.  Some residues are seen to be highly ordered in the native
state, while others are not.  We looked at the fastest residues whose
equilibrium level was at least 70\% ordered (bold residues in the
Tables).

The four fastest two-state residues are shown in Figure
\ref{fig:fast_residues}.  These four make key contacts between the
first hairpin and the helix.  Phenylalanine 30 and leucine 5 have a
strong hydrophobic interaction that secures the first strand of the
hairpin against the helix, while threonines 18 and 25 lock in the
second beta strand.  The fastest two-state residues are thus seen to
be important in forming the kinetic intermediate.  All residues
involved in intermediate formation are naturally found to be
two-state, because formation of the intermediate is a purely two-state
process.

The three-state residues are ones whose sidechain ordering cannot
proceed normally until the intermediate has formed.  They exhibit a
lag phase as seen in Figure \ref{fig:igd_traces} while the
intermediate forms.  In Figure \ref{fig:three_state} we show the three
fastest three-state residues of Protein G which are significantly
ordered at equilibrium: valines 6 and 54 and phenylalanine 52.
Interestingly, these three residues all have the same rate of
relaxation, suggesting that they become ordered together.  All three
are involved in bringing beta-strand 4 in hairpin 2 into contact with
the rest of the protein.  Valine 6 establishes contacts between
beta-strands 1 and 4.  Valine 54 makes contacts with valine 39
(located at the C-terminus of the helix) which hold the end of hairpin
2 against the helix. Phenyalanine 52 makes hydrophobic contacts with
tyrosine 45, stabilizing hairpin 2 internally, while also making
contacts with the helix.

The data obtained from the all-atom simulation is in good qualitative
agreement with our lattice simulation.  There is a wide distribution
of residue relaxation rates, with the fast residues located in
topologically important positions.  The same mechanism seems to be at
work here: key organizing residues form quickly holding the overall
structure together, while all other residues relax more slowly toward
equilibrium via fluctuations around the native fold.  On the lattice
we found strong overlap between the fast residues and the nucleus
residues which organized the backbone transition state.  In order to
make a similar comparison in the all-atom model, we proceeded to
characterize its transition state ensemble.

Because the transition state ensemble lies at the top of the folding
free energy landscape, its conformations are characterized by a 0.5
probability of folding ($p_{\mathrm{fold}}$) during a tiny fraction of
the entire folding time (``commitment
time''~\cite{kn:ropes_geissler}).  Assuming a commitment time
corresponding to 0.005\% of a full folding run, we calculated the
$p_{\mathrm{fold}}$ of approximately 5 structures per trajectory.  A
histogram of contacts (Figure \ref{fig:residue_contacts}) made by each
residue for various $p_{\mathrm{fold}}$ ensembles reveals that
phenylalanine 52 is the most important residue for the final
intermediate $\rightarrow$ native folding step.  Its energy
contribution, which is proportional to the number of contacts it
makes, appears to grow as the ensemble $p_{\mathrm{fold}}$ approaches
one.  Though less pronounced, similar increases were seen for Y3, K4,
L5, V6, A23, E27, F30, W43, Y45, K50, and V54.  When individual
residue-residue contacts are histogrammed (Figure
\ref{fig:residue_contacts_detail}), it is clear that only a handful of
over 1500 possible contacts are important for stabilizing the
transition state ensemble.  These special contacts bring two hairpin 2
residues (F52 and V54) in contact with hairpin 1 (Y3, L5, V6) and
helix residues (E27 and F30).  Because of the non-local and specific
nature of these contacts, folding in this model appears to be
consistent with that proposed under the theory of specific
nucleation~\cite{kn:specific_nucleus,kn:fersht_nucleation,kn:molecular_collapse,kn:eugene_review,kn:new_view}.
Detailed comparison of these results with experimental data will be
presented elsewhere (JS, EIS, manuscript in preparation).

It is clear that the nucleus characterizing the transition state
ensemble under our all-atom model of folding is nicely identified by
the time-series analysis of sidechain dynamics.  The three fastest
three-state residues - V6, F52, and V54 - coincide with those which
are most indicative of progress along the $p_{\mathrm{fold}}$
hypersurface.  Although structures with $p_{\mathrm{fold}} \approx 1$
will rapidly attain native-like backbone topologies, energies will
reach equilibrium values very slowly, requiring simulations extending
beyond the $2 \times 10^{9}$ cutoff we have used here.  This is
because a fairly significant amount of energy is contributed by
sidechain-sidechain interactions, and the correct packing of
sidechains is significantly slower once the collapse transition has
occurred.  In our previous study of crambin, we observed a similar
phenomenon (which we referred to as the ``sidechain-packing trap'';
see Figure 6E in reference~\cite{kn:crambin_folding}), where the
folding of the backbone occurred on a faster timescale than that by
which the native energy was fully attained.  The current
$p_{\mathrm{fold}}$ analysis demonstrates that, in fact, not all
residues participate equally in the slow relaxation of conformations
with incorrect packing.  The nucleus residues (V6, F52, V54) have to
attain native packing relatively early as their energy contribution is
required to counterbalance the tremendous loss of backbone entropy
upon collapse to a native-like conformation.

Finally, we also note a striking similarity between the thermodynamic
data of the lattice model presented here and crambin obtained from
all-atom simulations.  For crambin, we observed a rather unusual
departure from a simple two state model when fitting the equilibrium
energy against temperature.  As temperature was lowered below the
transition point, the decrease in energy was perfectly linear with
temperature.  As shown in Figure~\ref{fig:thermo} (particularly for
the 48-mer), as the number of sidechain states is increased, the same
linear relation between energy and temperature is observed.  This
suggests that sidechain degrees of freedom lead to a noticeable
contribution to the heat capacity, which dominates the thermodynamic
behavior at low temperatures.

\section*{Discussion}
Extracting information about the dynamics of individual sidechains is
relatively easy in computational studies and veritably challenging in
experiments.  There are several difficulties to overcome in
experiments.  First, specific probes that measure properties about a
single residue are scarce: tryptophan can be probed by fluorescence,
while cystein can be probed by thiol-disulfide exchange.  While
dynamic NMR techniques can in principle report on many residues
simultaneously, their application requires very slow folding
reactions.  Hydrogen exchange experiments~\cite{kn:hdx_englander} can
report on the protection of individual backbone amide groups, but
backbone protection factors do not directly measure sidechain
mobility.  Second, it is desirable to have probes in several different
parts of a structure in order to measure the distribution of sidechain
rate constants over the whole fold.  This, again, is in principle
possible but usually requires introducing sequence mutations (adding a
tryptophan or cystein).  Results must therefore be handled with care
because the structure and folding pathways may be altered in subtle
ways from sequence to sequence.  Finally, the presence of kinetic
intermediates in the folding of many proteins complicates analysis
considerably.

Several recent studies have attacked the sidechain dynamics question
using a variety of techniques.  Staniforth and
coworkers\cite{kn:waltho_immobilization} used a form of cystatin in
which disulfide bonds were reduced, thus creating a molten globule
whose compactness and unfolding properties were similar to folded wild
type, but whose sidechain mobility was significantly increased.  The
size of the rate-limiting barrier for folding of the two forms was
measured and found to be similar.  The authors conclude that since the
reduced and wild type forms differ mainly in sidechain mobility, while
the barrier height for folding is the same, the immobilization of
sidechains occurs after the major folding transition in wild type
cystatin.  Additional experiments on cystatin are probably needed in
order to completely solidify the argument.  Specifically, the
connection between fluoresence quenching upon folding and full
sidechain immobilization in wild type cystatin has not been made;
thus, any conclusions about sidechain immobilization rest on the
assumption that nativity of tryptophan fluoresence gives information
about sidechain dynamics across the entire core.

Ha and Loh\cite{kn:loh_apomyoglobin} introduced cystein mutations in
several key places in apomyoglobin and, using pulsed thiol-disulfide
exchange at different times during the folding reaction, measured the
progression of side chain ordering at each site.  They found that
certain locations, stabilizing the fast-forming folding intermediate,
were as well-packed as native protein long before folding was
complete.  It would be interesting to obtain similar site-specific
time courses for other positions distributed across the protein and to
see whether positions that become ordered in the post-intermediate
step exhibit a distribution of relaxation times. 

In an elegant series of experiments using time-resolved fluoresence
anisotropy measurements, Sridevi and
coworkers\cite{kn:sridevi_barstar} demonstrated that barstar's
tryptophan 53 becomes fully ordered approximately 8 times faster than
the rate of the slow folding reaction of the protein.  By observing
fluoresence lifetime decay, they could watch the initially evenly
populated rotamers of tryptophan reach nativity in which one rotamer
is 88\% populated.  The authors suggest that rapid relaxation of
tryptophan indicates the existence of an intermediate during the slow
folding of barstar.  It is not clear, however, that this must be the
case.  An alternate explanation is that there exists a significant
spread among sidechain relaxation rates within a single folding
reaction.

Our work demonstrates that the presence of sidechain degrees of
freedom leads to a wide distribution of residue relaxation rates, even
within two-state cooperative folding reactions.
Figure~\ref{fig:cartoon} gives a schematic overview of the relaxation
mechanism we observed.  Both in lattice and in all-atom simulations,
we found a small number of residues becoming fully ordered much faster
than the rest of the protein.  This observation is consistent with the
nucleation-condensation view of protein folding in which the major
transition state of the folding reaction involves a few residues
reaching their native conformation.  Importantly, in our simulations,
we find that these nucleating residues are not only in correct spatial
geometry with respect to each other's centers of mass, but
additionally their native rotamer has been singled out and practically
frozen.  Once nucleation has occured, the native chain topology is
strongly stabilized and certain measures, such as compactness and
perhaps fluoresence, might indicate that the reaction is complete, and
equilibrium has been reached (see Figure~\ref{fig:cartoon} after
nucleation barrier).  This, however, is not the case as there exist
many sidechains that have become partially ordered, yet have still not
reached equilibrium.  Because the nucleating residues have frozen and
are rigid, and many other partially ordered residues are significantly
stabilizing the fold, the non-equilibrated sidechains are not able to
convert easily to their native rotamer.  They remain in a non-native
state until a backbone fluctuation momentarily allows them to
interconvert.  The presence of backbone breathing motions in protein
globules may therefore be useful not only for function, as has been
suggested before~\cite{kn:alexandrov,kn:petsko_flexibility}, but also
in order to allow sidechain equilibrium to be achieved in a reasonable
amount of time.

In simulations, a kinetic intermediate is very easily observed as a
set of conformations which appears as a plateau within some range of
energies during many folding
trajectories~\cite{kn:lattice_intermediates,kn:crambin_folding}.  In
the Protein G simulations, the major folding pathway consisted of
formation of hairpin 1 and the helix followed by formation of hairpin
2, while in the minor pathway the intermediate consisted of hairpin 2
and the helix.  Each run proceeded through one and only one
intermediate, and the major pathway was observed in twice as many
runs.  It appears that the pathways observed in our simulation are
consistent with available experimental data on Protein G - this will
be discussed at length elsewhere, and does not bear significantly on
the present work.

The existence of a folding intermediate in our simulation of Protein
G, while complicating our analysis somewhat, has one important
advantage: we are able to see that the kinetics of only half of the
sidechains are sensitive to the presence of the intermediate; the
other residues exhibit single-exponential relaxation.  In other words,
a kinetic intermediate can be completely invisible if the wrong
position is used to probe folding.  We observed a distribution of
residue relaxation rates for both the pre- and post-intermediate
steps.  Each of these steps is a purely two-state process as seen by
the abrupt jump in rms deviation and energy.  It appears, then, that a
few key residues reach nativity faster than all others and propel the
chain through its transition state.  Further relaxation after the
major event via backbone fluctuations yields a distribution of rates
over the fold, the exact nature of which is governed by the extent of
backbone mobility at each position in the ensemble.

At first glance this observation runs contrary to the belief that in
two-state transitions all parts of the structure must reach nativity
at the same rate.  The argument goes that if structure is obtained
gradually, with some parts folding faster than others, then there are
many distinct ensembles of states for the chain to traverse.  To
dispell this fear, it is crucial to note that the core residues which
are observed in simulation to freeze fastest also happen to be in key
organizing positions.  The ensemble of conformations consistent with
their freezing is highly native and therefore extremely small compared
with the ensemble of unfolded conformations.  The major transition of
protein folding occurs between these two ensembles and is a two-state
transition in simulation as in reality.  The entire molecule does not,
however, necessarily reach equilibrium concomitantly with this barrier
crossing.  There can be many other smaller barriers associated with
backbone fluctuations which need to be crossed in order for all
sidechains to reach equilibrium (see Figure~\ref{fig:cartoon}).  

It is important to note that temperature plays a key role in making
sidechain relaxation possible in a reasonable amount of time.  At low
temperatures backbone fluctuations are small and sidechain relaxation
is a very noticeable and very long process, as seen in Figure
\ref{fig:energy_relax}.  At optimal folding temperature, however, the
energy of the post-nucleation ensemble is very close to its
equilibrium value.  Sidechain relaxation is still very slow, following
stretched-exponential kinetics, but the product of the major
transition is significantly closer energetically (and therefore
structurally) to equilibrium.  This suggests that under optimal
conditions, the slow sidechain packing process may not be
physiologically relevant because the ensemble of folded yet
unequilibrated molecules is structurally close enough to the native
ensemble that it may exhibit similar amounts of protection from
proteolysis.  The relatively small gap between mispacked and native
molecules at these temperatures suggests that relevant experiments
must be sensitive enough to detect such differences.

Since we could not observe full equilibration in the all-atom
simulation, we return to the lattice simulations in order to discuss
the relevant post-nucleation processes which establish equilibrium.
In lattice simulations we found that the dynamics during short times
is reminiscent of the classic nucleation mechanism that has been
observed before~\cite{kn:specific_nucleus}. Due to the existence of
sidechain states, the nucleation-organized backbone does not reach
equilibrium immediately.  At long times the system can end up in traps
which require some degree of backbone motion to allow sidechains to
interconvert.  This suggests that perhaps the energy landscape after
the native fold has been acquired consists of a series of barriers,
associated with backbone fluctuations, which must be crossed.  As the
system traverses these barriers it moves to lower and lower energies.
If one assumes that the transition states for these barriers are
largely very similar, the resulting relaxation process can be shown to
be logarithmic in time~\cite{kn:rem_relaxation}.  In the lattice
simulations, we observed a highly stretched exponential relaxation at
long times.  Since the stretching exponent is very low at 0.09, the
time dependence of energy is essentially logarithmic.  The landscape
for slow sidechain equilibration thus seems to be one of increasingly
deeper wells, rather than a single cooperative transition to nativity.
This places sidechain relaxation within the set of phenomena that can
be characterized as a glass.  Classic lattice models without sidechain
states, however, do not exhibit a glass transition at any reasonable
temperature~\cite{kn:lattice_glass}.  Our lattice simulations indicate
that the presence of sidechain degrees of freedom may lead to glassy
relaxation, but further detailed characterization of the energy
landscape, as well as additional tests using more realistic models,
are required to solidify this claim.

Using both all-atom and lattice simulations, we have demonstrated that
full sidechain relaxation during protein folding can be a process
whose timescale is significantly slower than that of crossing the
major folding barrier.  While the major barrier is traversed via the
classic nucleation mechanism, we find that equilibrium is reached via
a set of smaller barrier crossings that correspond to backbone
fluctuations.  The heterogeneities inherent in protein structures give
rise to a distribution of sidechain relaxation times which can span up
to an order of magnitude.  A number of recent experiments are
consistent with our findings.  We hope that this work will spur
further dialogue between simulations and experiments to elucidate the
complex processes that bring sidechains to equilibrium.

\section*{Methods}
\textbf{Lattice Model with Internal Monomer States.} We use a standard
three-dimensional lattice model in which each monomer occupies a
single lattice site.  For the 27-mer simulations, a sequence was
designed to fold to a unique native conformation as described
in~\cite{kn:cumulant_design}.  For the 48-mer simulation, a
fast-folding sequence was obtained from a lattice protein evolution
study described in~\cite{kn:lattice_evolution}.  The standard
Miyazawa-Jernigan parameter set~\cite{kn:myazawa_jernigan} is used to
compute the energy of a conformation.  Two monomers are said to be in
contact if they are nearest neighbors on the lattice, and are not
sequence neighbors.  Additionally, each monomer has $n$ internal
states, where $n$ is a parameter of the model.  We present data for
27-mer folding with $n = 1, 2, 4, $ and $8$, and for 48-mer folding
with $n = 1, 2, $ and $4$.  The internal state of each monomer is
stored as a number from 0 to $n-1$.  The 0 state is the native state,
while the states $1$..$n-1$ are non-native.  If $n=1$ then all
monomers remain native throughout the simulation, and the model is
equivalent to the standard lattice model.  Non-native monomers do not
contribute to energy.  That is, two monomers in contact will
contribute to energy only if they are both in their native state, the
0 state.

The standard cubic lattice move-set~\cite{kn:lattice_moveset} is used
to evolve the backbone conformation, and a Metropolis
criterion~\cite{kn:metropolis} with temperature $T$ is used to
accept/reject moves.  In addition to backbone moves, the internal
states of the monomers must be evolved.  After each backbone move is
attempted, we attempt $n-1$ internal state moves.  At each such move,
a random monomer is chosen.  If the monomer is making more than $c$
contacts with other monomers, its internal state is not allowed to
change.  Otherwise, its internal state is randomly flipped to one of
the other $n-1$ states, the change in energy of the conformation is
computed, and the move is accepted/rejected based on the Metropolis
criterion.  The parameter $c$ can take the values 0 through 4.  When
$c = 4$, internal states can interchange freely and are not affected
by the conformation of the backbone.  If $c = 0$, internal states can
interchange only if the monomer makes no contacts.  In this study we
take $c = 0$ throughout, except in the control simulation (Figure
\ref{fig:kinetics}D) in which we use $c = 4$.  Folding simulations are
started from random backbone conformations generated by an infinite
temperature simulation.  The internal state of each monomer is
initialized randomly.

\textbf{All-Atom Protein Folding Simulations.}  The all-atom Monte
Carlo simulation previously described in~\cite{kn:crambin_folding} was
used to fold protein G (pdb code: 1IGD).  By representing all
sidechain and backbone heavy atoms as hard spheres, the protein was
simulated as a polymer with excluded volume interactions, where chain
crossings are strictly prohibited.  The energy of a conformation was
computed as $E = E_{G} + E_{h}$, where (1) the atom-atom G\={o} energy
$E_{G} =\sum C(A,B) \Delta(A,B)$, with $\Delta(A,B) = 1 $ if the
heavy atoms $A$ and $B$ were in contact and zero otherwise, $C(A,B)$
was -1 if $A$ and $B$ were in contact in the native conformation, 1 if
they were not, and $\infty$ if they were sterically clashed and (2)
the backbone hydrogen bonding energy $E_{h} = N_{h} h$, where $N_{h}$
corresponds to the number of amide N-carbonyl O pairs in contact.  $h$
was chosen to be -0.6 in order to match experimental stabilities of
the protein G helix and hairpins taken in isolation (JS and EIS,
manuscript in preparation).

The torsional move set ensures that canonical bond lengths and
geometries (including planar peptide bonds) are maintained throughout
the entire simulation.  Backbone and sidechain moves consisted of
concerted random rotations of backbone $\phi/\psi$ and sidechain
$\chi$ angles, respectively.  10 sidechain moves were completed for
each backbone move in order to allow sufficient relaxation of
sidechain geometries after a change in the backbone topology.

50 folding simulations were initiated from random coil conformations,
obtained by simulating the native state with only the excluded volume
interaction turned on.  The temperature was then quenched to T=1.575
and the chain was allowed to equilibrate for $2 \times 10^{9}$ MC
steps, where 1 MC step consisted of 1 backbone and 10 sidechain moves.
Given the experimentally measured transition temperature of $360 K$
and our simulation transition temperature of 1.95, T=1.575 corresponds
to an actual temperature of $\approx$ 290 K.

From the 50 trajectories of protein G, we estimated the probability to
fold~\cite{kn:pfold_du} ($p_{\mathrm{fold}}$) of conformations
observed just prior to reaching the native state, by counting the
number of times the native state was attained from the selected
conformation in 20 separate runs of $10 \times 10^6$ MC steps.

\textbf{Fitting of residue relaxation curves.} After collecting many
long runs, we averaged the internal sidechain state of each monomer at
each time step over all runs, assigning 1 if the residue was native,
and 0 otherwise.  For lattice simulations, 130 runs were used, and a
two-state exponential fit of the form $f(x) = a_0 + a_1 \exp(-a_2 t)$
was very good for all residues.  For all-atom simulations, 50 runs
were used, and averaging over runs was performed by assigning 1 to
each residue whose $\chi$-angles were all within $60^{\circ}$ of the
native angles, and 0 otherwise.  A value of 1 thus corresponded to
observing the native rotamer.  Fits to a three-state model were
performed as described in Results.  All fits were done using the
nonlinear least-squares Marquardt-Levenberg algorithm.

\section*{Acknowledgements}
We would like to thank Leonid Mirny and Phillip Geissler.  Research
was supported by NIH grant GM52126.

\bibliographystyle{myjmb} \bibliography{sc_dynamics}

\newpage
\section*{Figures}

\noindent
\underline{Figure 1} \\ Thermodynamics of 27-mer and 48-mer lattice
models.  Each point corresponds to an average of energy over a lattice
simulation started at the native state.  Each simulation was run for 3
$\times 10^8$ steps and energy was sampled every 3 $\times 10^5$
steps.  The correlation time of energy was found to be much less than
our sampling interval at all temperatures.  The number of internal
sidechain states for each model is indicated by the color and shape of
the points, as in the legend.  Fits to a two-state thermodynamic model
are given in solid lines colored to match the corresponding lattice
model that was used.  Parameters for these fits are given in Table
\ref{tb:thermo_fits}. \\
\noindent
\underline{Figure 2} \\ Kinetics of 27-mer lattice models.  In panels
\textbf{A}, \textbf{B}, and \textbf{C}, the kinetics of the models with
2, 4, and 8 sidechain states per monomer are shown.  Interconversion
of internal sidechain states in these models can take place only if a
given monomer is not in contact with any other monomer. The logarithm
of the mean first passage time (MFPT) to native energy is shown by
open circles, while the average time for reaching a stable native
backbone is given by squares.  By stable formation of the backbone we
mean that once formed, the backbone did not subsequently unfold by
more than 50\% before reaching the native energy.  In panel
\textbf{D}, two control models are shown: the n = 1 model, in which
each monomer has a single internal state, and thus corresponds to the
classic lattice model; and the n = 2 unhindered model, in which
monomers have 2 sidechain states, and these states can interconvert
freely, regardless of whether the given monomer is in contact with
others or not.  Each point was calculated over a set of between 100
and 200 runs, and error bars corresponding to 1.5 standard deviations
are indicated.

\noindent
\underline{Figure 3} \\ Kinetics of the 48-mer model with 4 internal
states per monomer.  Panel \textbf{A} indicates the MFPT for reaching
native energy by circles, and the average time to reach the stable
native backbone by squares.  Panel \textbf{B} shows the time to reach
the stable backbone with diamonds, and the amount of time to go from
the native backbone to the native energy with x-marks. \\

\noindent
\underline{Figure 4} \\ Average time traces for two representative
residues in the lattice 48-mer.  Residue 35 is a nucleus residue
exhibiting fast freezing, while residue 13 is a non-nucleus residue
with an average freezing rate.  The black line is the best
single-exponential fit to the data. \\

\noindent
\underline{Figure 5} \\ Histogram of residue relaxation rates for
48-mer with 4 internal states.  Histograms for low temperature (T =
7.4) and optimal folding temperature (T = 9.1) are shown.  Each
residue was assigned a value of 1 if it was in its native sidechain
state, and 0 otherwise, and these numbers were averaged at each time
step over 130 long runs.  Rates were calculated by fiting a single
exponential relaxation to the resulting native occupancy curves for
each residue.  At T = 7.4, runs of length 2 $\times 10^9$ were used;
at T = 9.1, run length was 2 $\times 10^8$.  The fast positions at
each temperature are labelled by numbers on the histograms.  Red
numbers correspond to positions which are more than 90\% ordered in
the native state, while green numbers are less than 90\% ordered.  \\

\noindent
\underline{Figure 6} \\ Lattice 48-mer structure colored by rate of
freezing at $T = 7.4$.  Nucleus positions, determined in
\cite{kn:eugene_evolution}, are indicated by large spheres.  Colors
range from white (slow-freezing) to blue (fast-freezing). \\

\noindent
\underline{Figure 7} \\ Relaxation of energy to equilibrium.  By
averaging energy at each time step over 130 long runs, we obtained
energy relaxation curves at temperatures 7.4 and 9.1.  The same runs
were used in Figure~\ref{fig:rate_hists}.  The green curves in each
figure are the average energy obtained from simulations.  The red line
corresponds to the average energy at equilibrium, and was obtained for
each temperature by averaging over a long run started at the native
lowest energy state, as in Figure~\ref{fig:thermo}.  The dashed line
is a fit to a three-state exponential model (see Methods).  A fit
using a two-state exponential model yielded a nearly identical curve.
The arrows indicate the average time to reach the stable native
backbone at each temperature.  The solid curve in panel \textbf{B} is
a two-phase fit using a single exponential ($a_0 + a_1 exp(-a_2
t/10^8)$) for short times, and a stretched exponential ($-1092 + b_1
\exp (-b_2 t^{\alpha})$) for long times.  Parameters of these fits are
$a_0 = -838$, $a_1 = 663$, $a_2 = 235$, $b_1 = 9.5 \times 10^5$, $b_2
= 12.1$, and $\alpha = 0.087$.  The value of $-1092$ for the equilibrium
energy corresponds to the red line.  \\

\noindent
\underline{Figure 8} \\ Average time traces for two representative
residues in Protein G.  P30 and P52 are typical two-state and
three-state residues, respectively.  The black line is the best fit as
described in Tables \ref{tb:igd_fits2} and \ref{tb:igd_fits3}. \\

\noindent
\underline{Figure 9} \\ Protein G residues exhibiting fastest
two-state relaxation to a highly ordered state.  The four fastest
residues whose relaxation curves fit well to a two-state kinetic
model, and whose equilibrium conformation is at least 70\% ordered are
shown in pink.  These residues occupy key positions in the
major-pathway intermediate that is seen in all-atom simulations of
Protein G folding.  The helical residue F30 is lodged between L5 of
beta-strand 1, and T18 of beta-strand 2, thus organizing the entire
structure of the intermediate which consists of hairpin 1 and the
helix.  T25 makes contacts at the hairpin-helix turn. \\

\noindent
\underline{Figure 10}\\ Protein G residues exhibiting fastest
three-state relaxation to highly ordered state.  Residues V6, F52, and
V54, shown in pink, exhibited fastest three-state relaxation, and
remained highly ordered at equilibrium.  All three are important
post-intermediate positions: F52 and V54 secure strand 4 of hairpin 2
to the helix, while V6 makes contacts between the two hairpins.   \\

\noindent
\underline{Figure 11}\\ Dependence of residue nativity upon
$p_{\mathrm{fold}}$.  Conformations were binned according to their
$p_{\mathrm{fold}}$ values, and the average change in number of
contacts, with respect to the $p_{\mathrm{fold}} = 0$ state, is
plotted for each residue.  Each curve corresponds to an average over
all conformations within the given range of $p_{\mathrm{fold}}$
values. \\

\noindent
\underline{Figure 12} \\ Dependence of specific contacts upon
$p_{\mathrm{fold}}$.  As in Figure \ref{fig:residue_contacts},
conformations were binned according to their $p_{\mathrm{fold}}$
values.  For each $p_{\mathrm{fold}}$ range, the average change (with
respect to the $p_{\mathrm{fold}} = 0$ state) in number of atom-atom
contacts between all pairs of residues that make native contacts, is
plotted for each pair of residues.  Residues pairs are arbitrarily
ordered in a linear fashion along the x-axis.  \\
\newpage

\noindent
\underline{Figure 13} \\ Schematic diagram of barriers and their
significance during the folding reaction.  The first barrier
corresponds to the nucleation event which organizes the backbone
topology.  Associated with this barrier is the freezing of a small
group of residues - the nucleus - into their native sidechain states
(blue dots).  Other residues may still be partially disordered (red
dots).  The disordered residues become increasingly native-like via
barriers corresponding to backbone fluctuations which momentarily free
a few residues (see small arrows), and allow their sidechain states to
interchange.  Barriers become higher as chain approaches equilibrium.
\\
\newpage

\noindent
\begin{table}[h]
\begin{tabular}{ c c c c }
n & $a_0$ & $a_1$ & $a_2$ \\ \hline
\multicolumn{4}{c}{27-mer} \\ 
1 & -111 & 518 & 16.9 \\
2 & -71.7 & 500 & 23.4 \\
4 & -44.2 & 433 & 26.3 \\
8 & -18.2 & 320 & 24.0 \\ \\
\multicolumn{4}{c}{48-mer} \\ 
1 & -82 & 263 & 13.8 \\
2 & -3.6 & 222 & 16.0 \\
4 & 38 & 151 & 14.2 \\ \hline
\end{tabular}
\caption[]{Two-state Fits to Thermodynamic Data.  Thermodynamics shown
in Figure \ref{fig:thermo} was fit using the form $f(x) = a_3 + (a_0 -
a_3)\exp(a_2 - a_1/T)/(1 + \exp(a_2 - a_1/T))$, where $a_3$ is the
native state energy for each model.  For the 27-mer, $a_3 = -1219$;
for the 48-mer, $a_3 = -1361$.}\label{tb:thermo_fits}
\end{table} 
\newpage

\noindent
\begin{table}[h]
\begin{tabular}{ l c c c c c c }
\# &  \multicolumn{1}{c}{$a$} & \multicolumn{1}{c}{$b$} & \multicolumn{1}{c}{$c$} & \multicolumn{1}{c}{$d$} & \multicolumn{1}{c}{err $a$} & \multicolumn{1}{c}{err $b$} \\ \hline 
13K & 0.174 & - & 0.064 & 0.094 & $\pm$ 0.009 & - \\
49T & 0.190 & - & 0.150 & 0.483 & 0.008 & - \\
\textbf{37N} & 0.756 & - & 0.603 & 0.826 & 0.003 & - \\
36D & 0.779 & - & 0.168 & 0.326 & 0.01 & - \\
15E & 0.791 & - & 0.302 & 0.368 & 0.005 & - \\
\textbf{7I} & 0.941 & - & 0.676 & 0.965 & 0.003 & - \\
\textbf{21V} & 1.027 & - & 0.357 & 0.739 & 0.007 & - \\
\textbf{0V} & 1.035 & - & 0.379 & 0.813 & 0.01 & - \\
24E & 1.159 & - & 0.011 & 0.060 & 0.2 & - \\
32Q & 1.159 & - & 0.110 & 0.198 & 0.02 & - \\
27E & 1.283 & - & 0.442 & 0.539 & 0.008 & - \\
\textbf{33Y} & 1.308 & - & 0.810 & 0.991 & 0.003 & - \\
31K & 1.365 & - & 0.444 & 0.501 & 0.007 & - \\
\textbf{16T} & 1.504 & - & 0.606 & 0.958 & 0.005 & - \\
\textbf{3Y} & 1.581 & - & 0.779 & 1.000 & 0.003 & - \\
\textbf{17T} & 1.626 & - & 0.544 & 0.884 & 0.006 & - \\
\textbf{5L} & 1.724 & - & 0.636 & 1.000 & 0.004 & - \\
22D & 1.809 & - & 0.423 & 0.554 & 0.01 & - \\ 
28K & 1.882 & - & 0.026 & 0.057 & 0.00 & - \\
\textbf{18T} & 1.891 & - & 0.632 & 0.991 & 0.005 & - \\
\textbf{25T} & 1.931 & - & 0.596 & 0.985 & 0.005 & - \\
\textbf{30F} & 2.068 & - & 0.723 & 1.000 & 0.004 & - \\ \hline

\end{tabular}
\caption[]{Two-State Residues and Fits for Protein G.  Individual residue relaxation curves were initially fit to the following three-state kinetic model: $f(x) = d + c(a/(b-a))exp(-b x/10^9) - c(b/(b-a)) exp(-a x/10^9)$.  The parameter $d$, corresponding to fully equilibrated value of residue ordering, was obtained from long equilibrium simulation, and was not varied in the fitting process.  Standard non-linear fitting was used to calculate $a$, $b$, and $c$.  The residues listed in this table had one rate constant that was at least an order of magnitude faster than the other.  The fits listed are therefore essentially two-state fits, and we report only the relevant slow rate constant.  The three-state model was used in order to determine which residues were markedly two-state, and which ones were not.  Asymptotic error on parameter $a$ is listed as well.  The table is sorted by the rate constant $a$.}\label{tb:igd_fits2}
\end{table} 

\newpage
\noindent
\begin{table}[h]
\begin{tabular}{ l c c c c c c }
\# &  \multicolumn{1}{c}{$a$} & \multicolumn{1}{c}{$b$} & \multicolumn{1}{c}{$c$} & \multicolumn{1}{c}{$d$} & \multicolumn{1}{c}{err $a$} & \multicolumn{1}{c}{err $b$} \\ \hline 
\textbf{12L} & 0.279 & 1.974 & 0.601 & 0.842 & $\pm$ 0.008 & $\pm$ 0.1 \\
8N & 0.328 & 2.538 & 0.136 & 0.254 & 0.02 & 0.6 \\
46D & 0.464 & 2.524 & 0.058 & 0.191 & 0.07 & 1.0 \\
\textbf{55T} & 0.710 & 2.158 & 0.384 & 0.710 & 0.02 & 0.2 \\
35N & 0.722 & 3.301 & 0.034 & 0.187 & 0.1 & 1.8 \\
47D & 0.834 & 1.387 & 0.130 & 0.294 & 0.2 & 0.5 \\
\textbf{1T} & 0.859 & 5.414 & 0.426 & 0.780 & 0.01 & 0.3 \\
4K & 1.018 & 9.252 & 0.178 & 0.210 & 0.01 & 1.1 \\
10K & 1.02 & 1.02 & 0.027 & 0.053 & 0.03 & 0.03 \\
42V & 1.127 & 1.127 & 0.180 & 0.508 & 0.01 & 0.01 \\
50K & 1.205 & 1.205 & 0.369 & 0.386 & 0.004 & 0.004 \\
56E & 1.319 & 1.319 & 0.610 & 0.666 & 0.003 & 0.003 \\
44T & 1.361 & 1.361 & 0.341 & 0.689 & 0.006 & 0.006 \\
\textbf{43W} & 1.623 & 3.455 & 0.750 & 1.000 & 0.01 & 0.06 \\
\textbf{39V} & 1.719 & 1.719 & 0.567 & 0.984 & 0.004 & 0.004 \\
\textbf{45Y} & 1.734 & 1.734 & 0.883 & 0.999 & 0.002 & 0.002 \\
\textbf{53T} & 1.749 & 1.749 & 0.517 & 0.850 & 0.005 & 0.005 \\
\textbf{51T} & 2.020 & 2.020 & 0.510 & 0.841 & 0.006 & 0.006 \\
\textbf{54V} & 2.034 & 2.034 & 0.687 & 1.000 & 0.003 & 0.003 \\
\textbf{6V} & 2.035 & 2.035 & 0.642 & 0.966 & 0.004 & 0.004 \\
\textbf{52F} & 2.039 & 2.039 & 0.858 & 1.000 & 0.002 & 0.002 \\
19K & 2.111 & 3.089 & 0.049 & 0.068 & 0.4 & 0.9 \\
40D & 2.198 & 2.198 & 0.037 & 0.170 & 0.07 & 0.07 \\
2T & 2.923 & 2.923 & 0.285 & 0.628 & 0.02 & 0.02 \\
\hline
\end{tabular}
\caption[]{Three-State Residues and Fits for Protein G.  Fits were performed as in Table \ref{tb:igd_fits2}.  The two rate constant obtained for the residues listed here were within one order of magnitude of each other.  The table is sorted by the slower of the two rate constants, which is arbitrarily designated to be parameter $a$.}\label{tb:igd_fits3}
\end{table} 

\newpage
\pagestyle{empty}
\newpage
\begin{center}
\epsfig{file=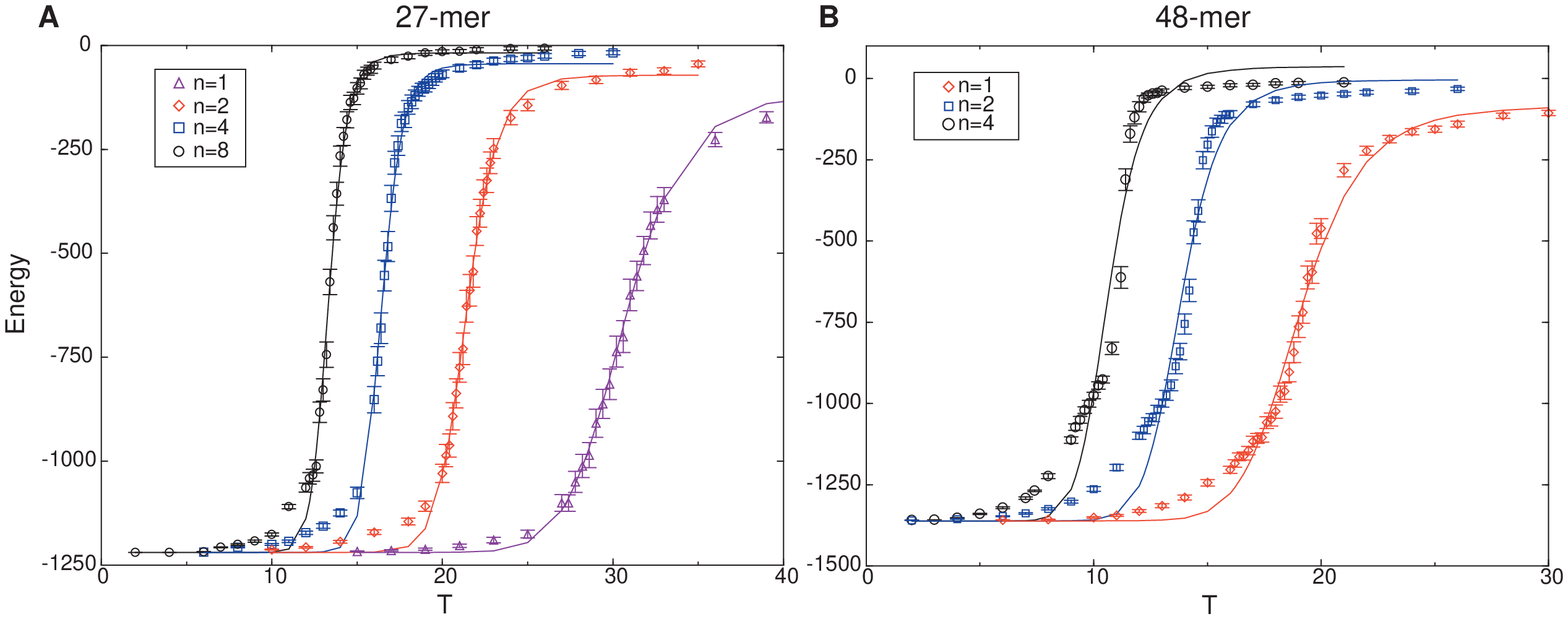,angle=90,height=\textheight}
\end{center}
\begin{center}
\epsfig{file=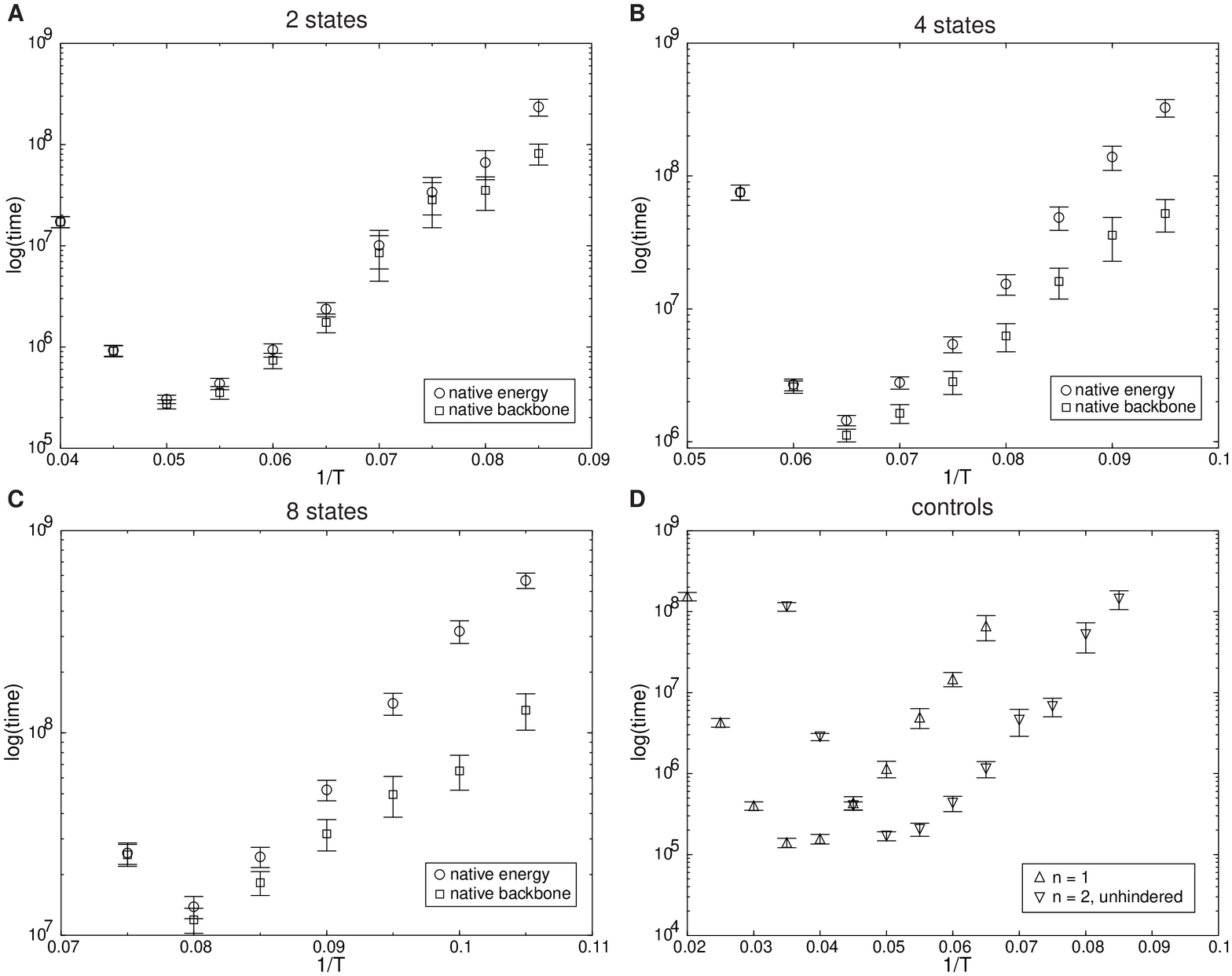,angle=90,height=\textheight}
\end{center}
\begin{center}
\epsfig{file=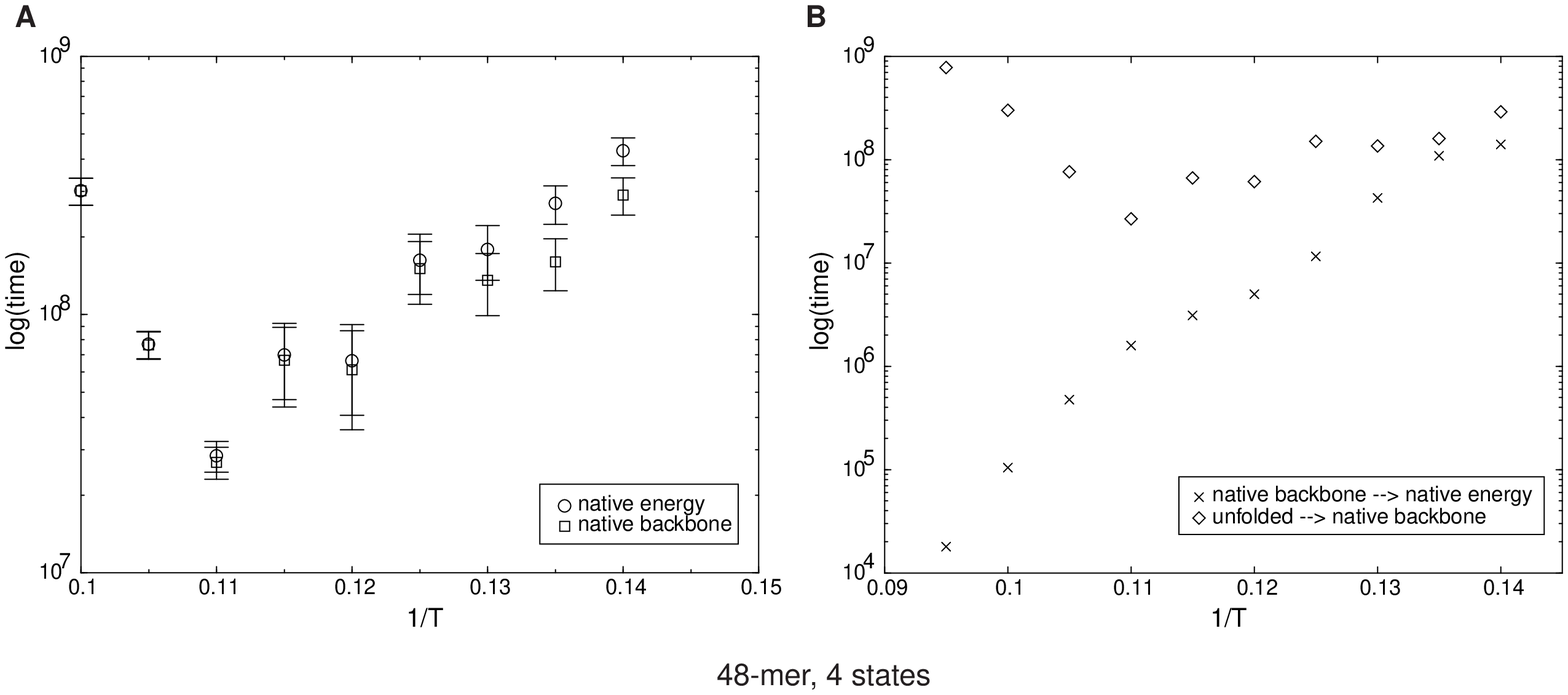,angle=90,height=\textheight}
\end{center}
\begin{center}
\epsfig{file=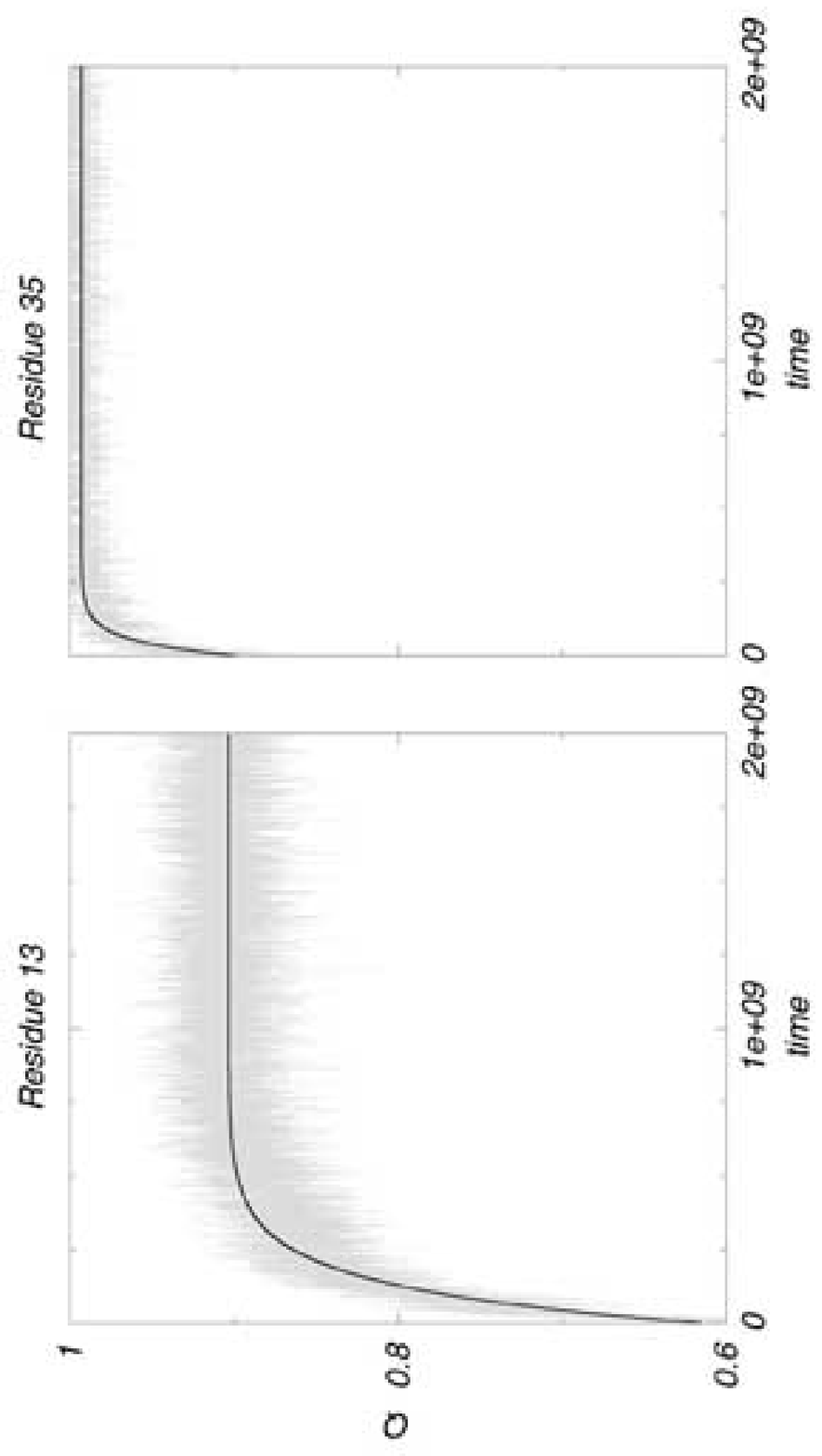,height=\textheight}
\end{center}
\begin{center}
\epsfig{file=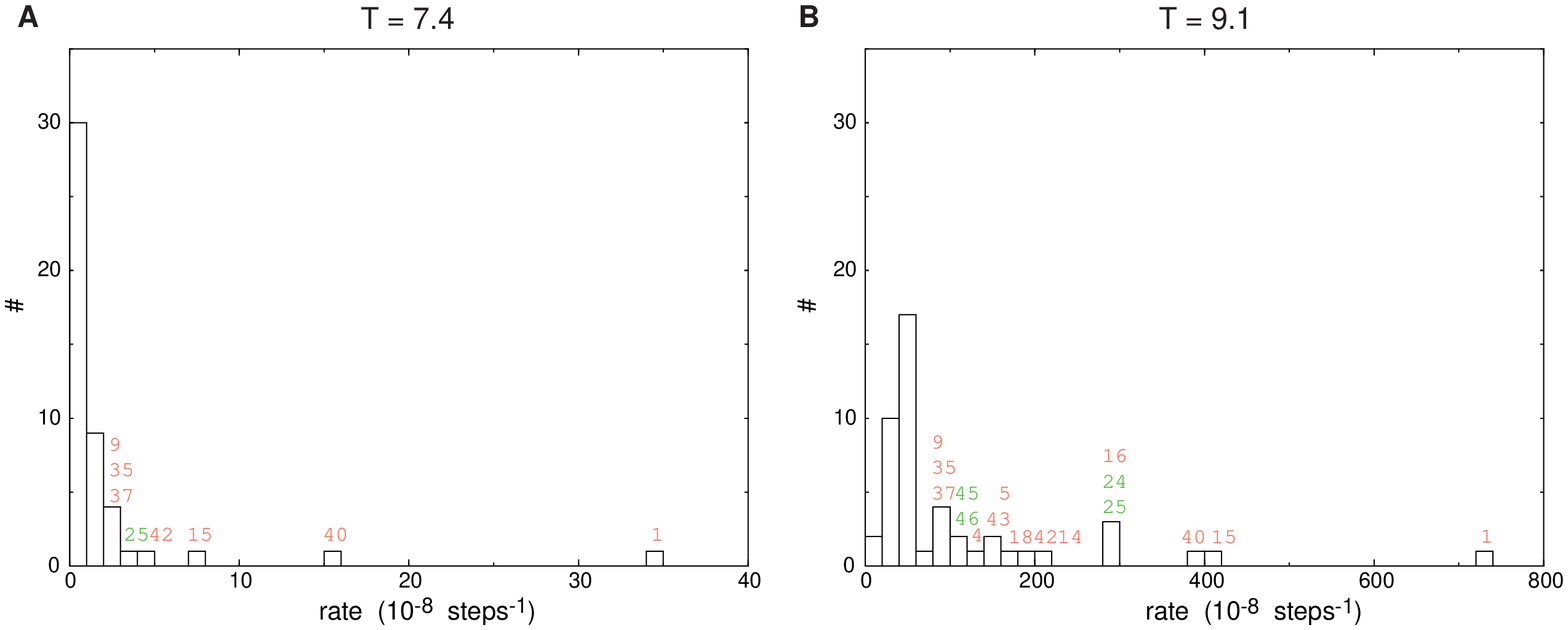,angle=90,height=\textheight}
\end{center}
\begin{center}
\epsfig{file=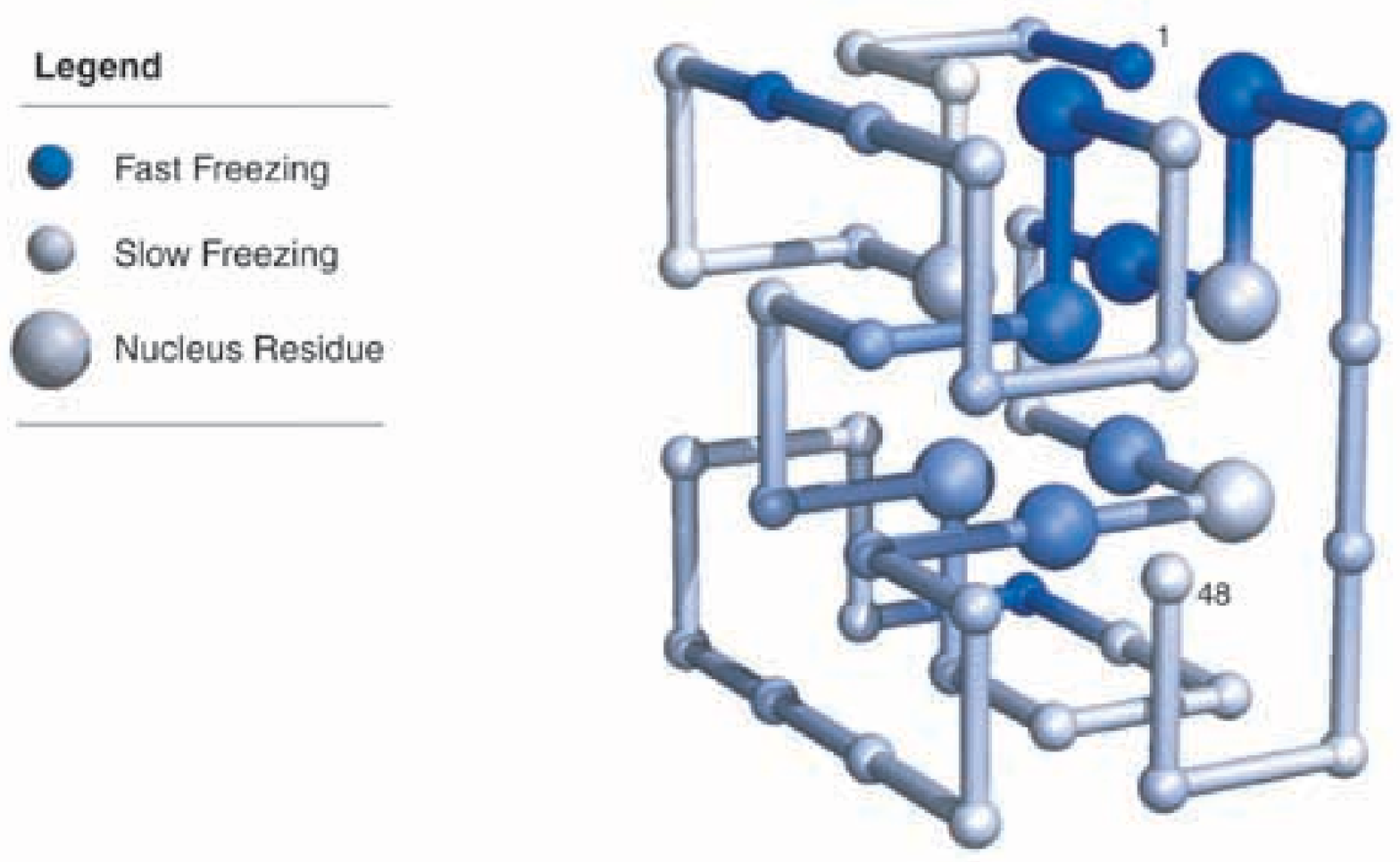,angle=90,height=\textheight}
\end{center}
\begin{center}
\epsfig{file=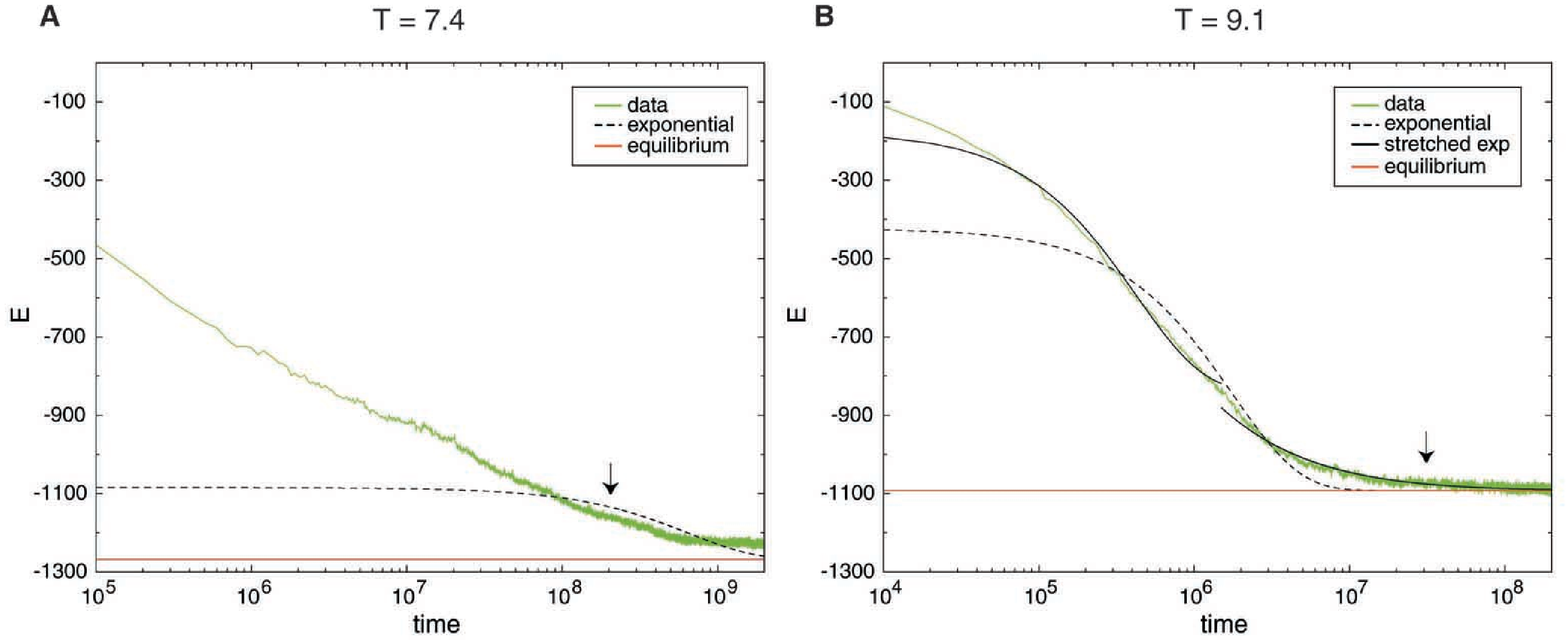,angle=90,height=\textheight}
\end{center}
\begin{center}
\epsfig{file=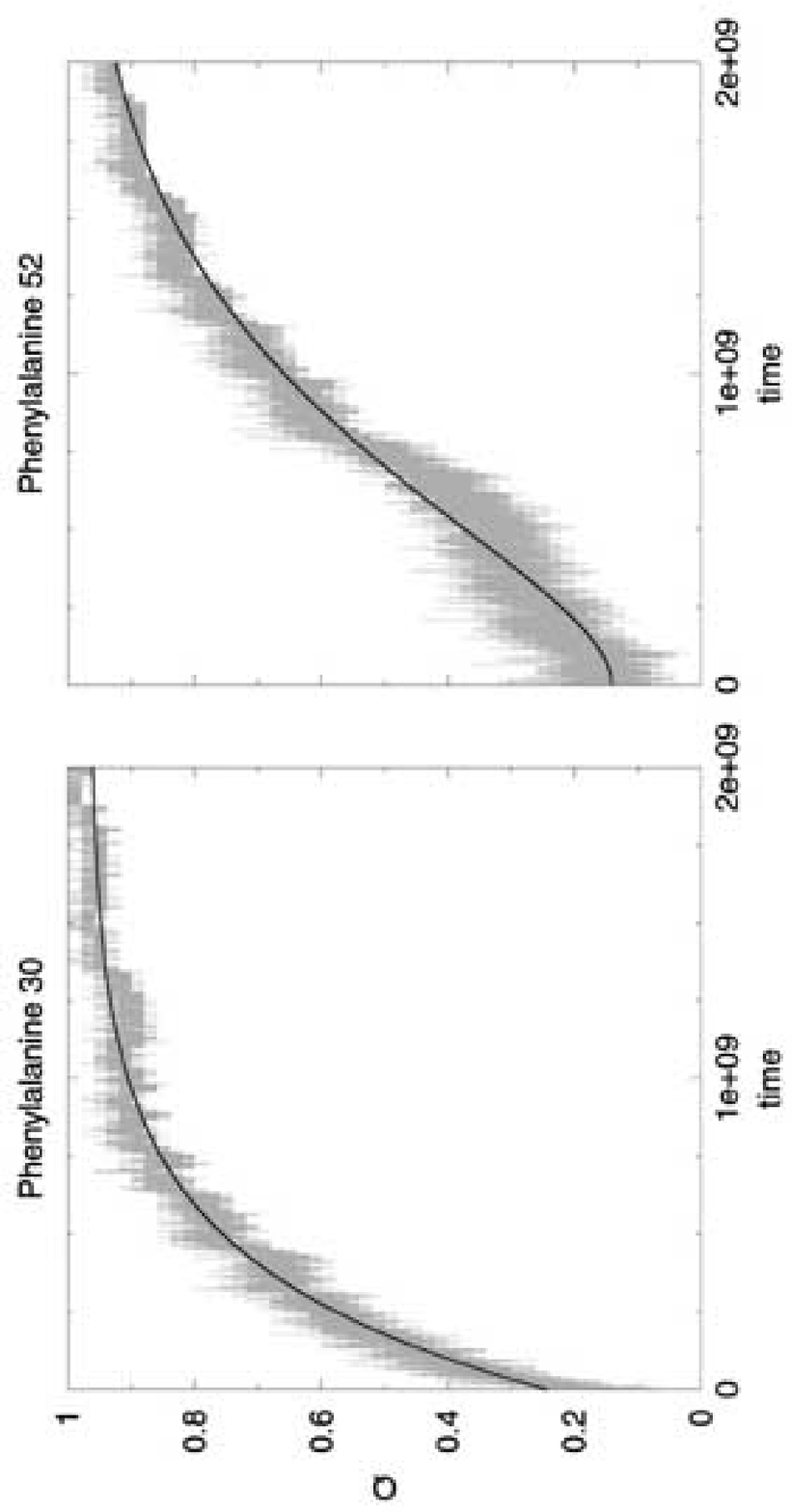,height=\textheight}
\end{center}
\begin{center}
\epsfig{file=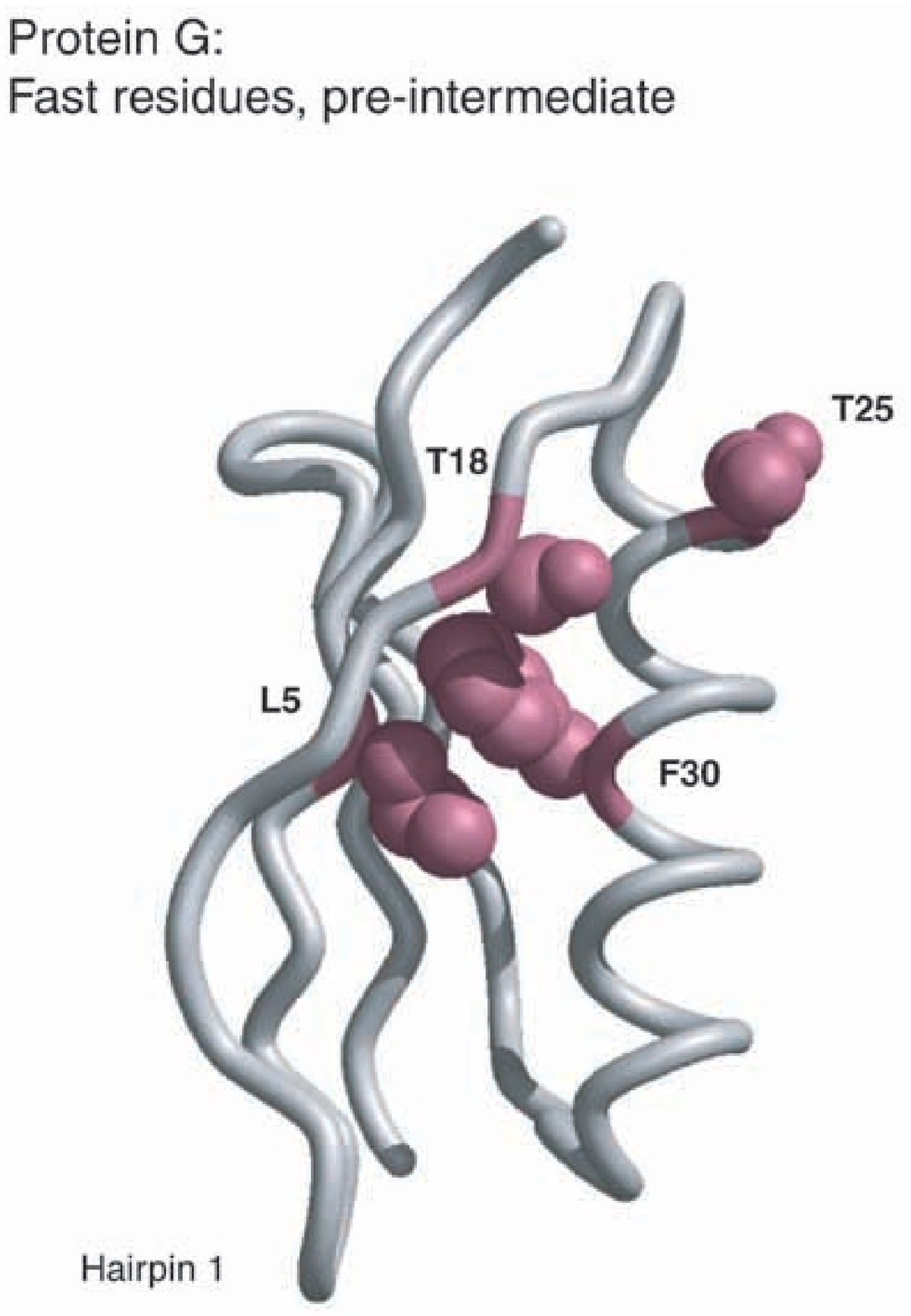}
\end{center}
\begin{center}
\epsfig{file=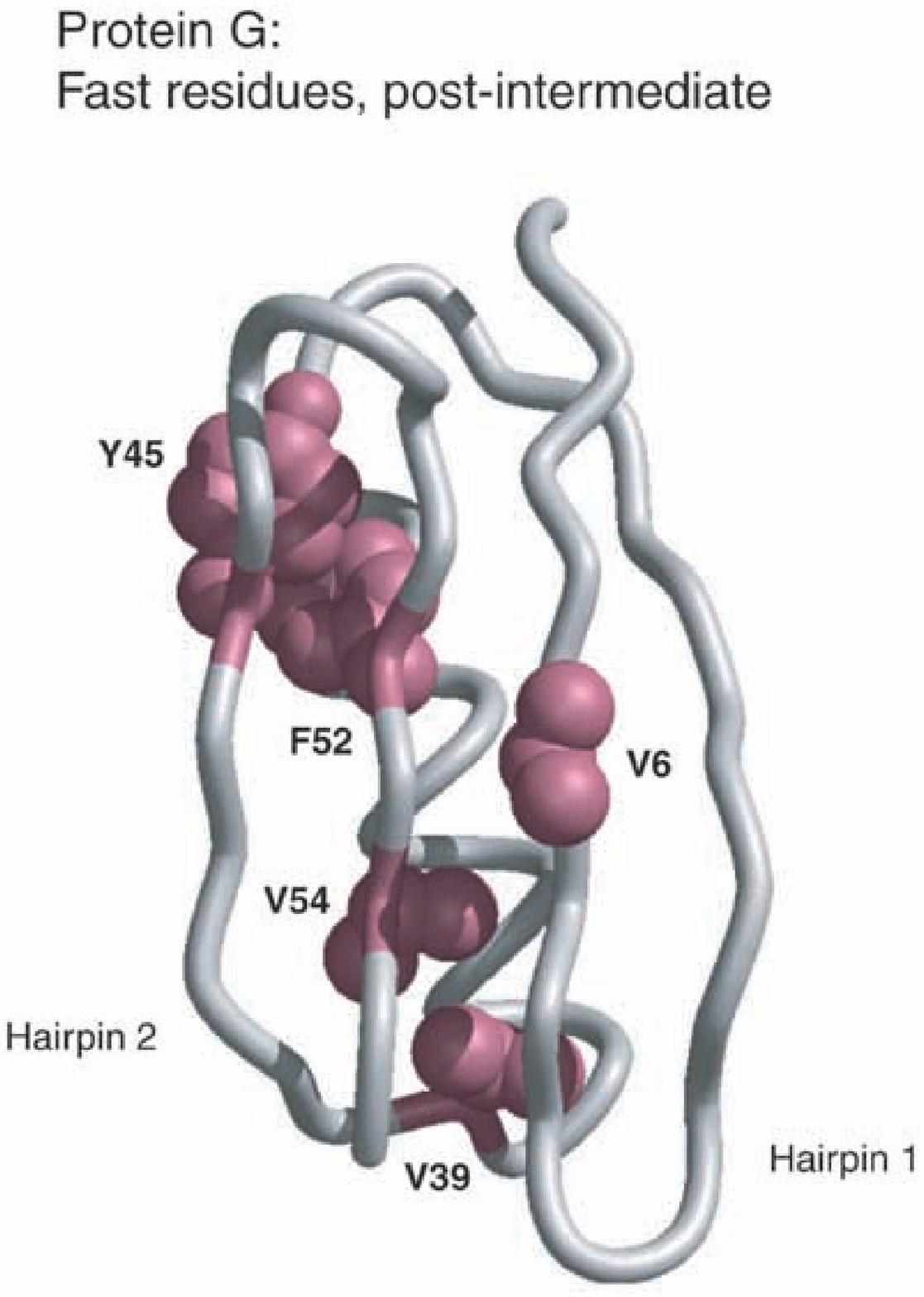}
\end{center}
\begin{center}
\epsfig{file=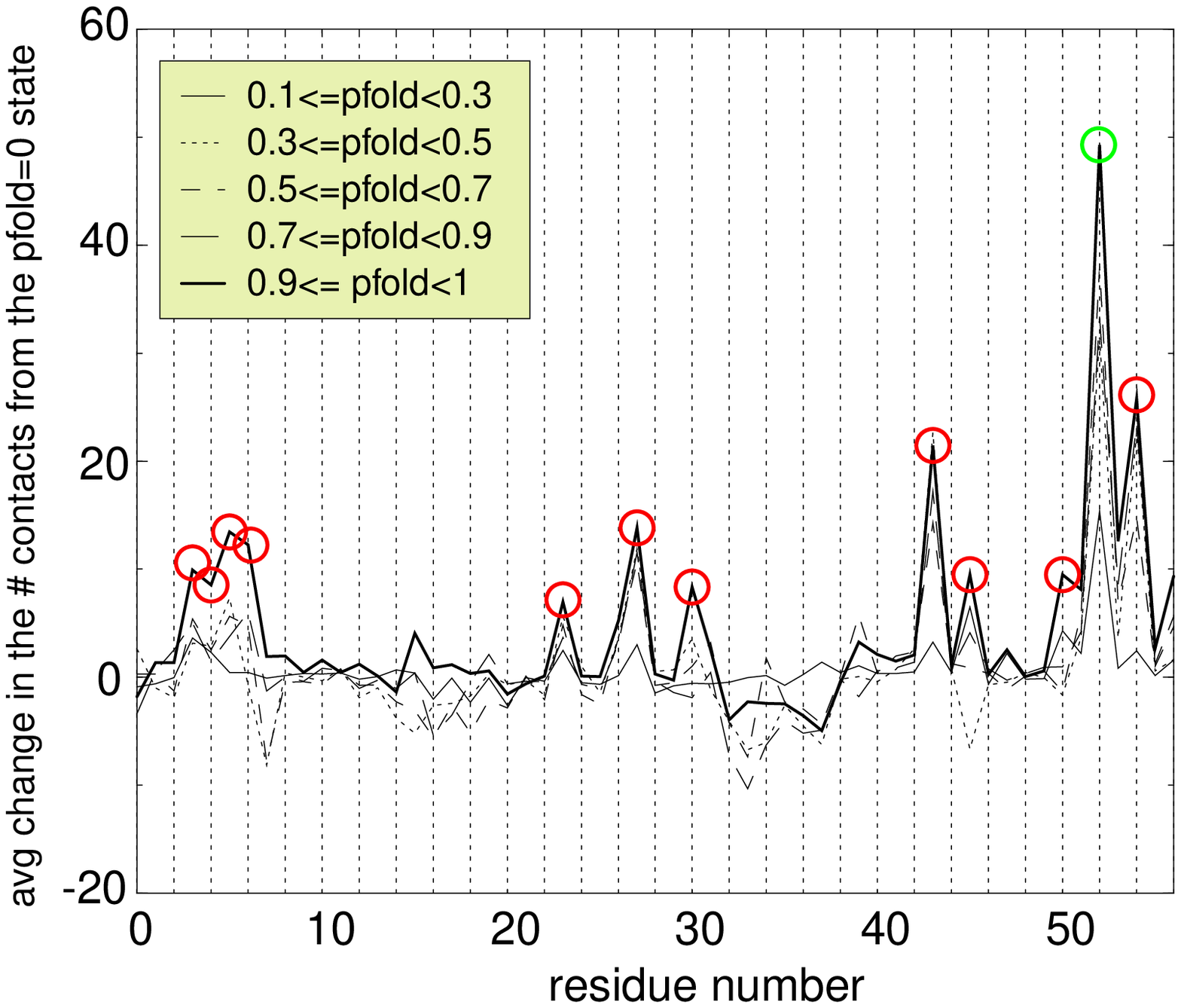,angle=90}
\end{center}
\begin{center}
\epsfig{file=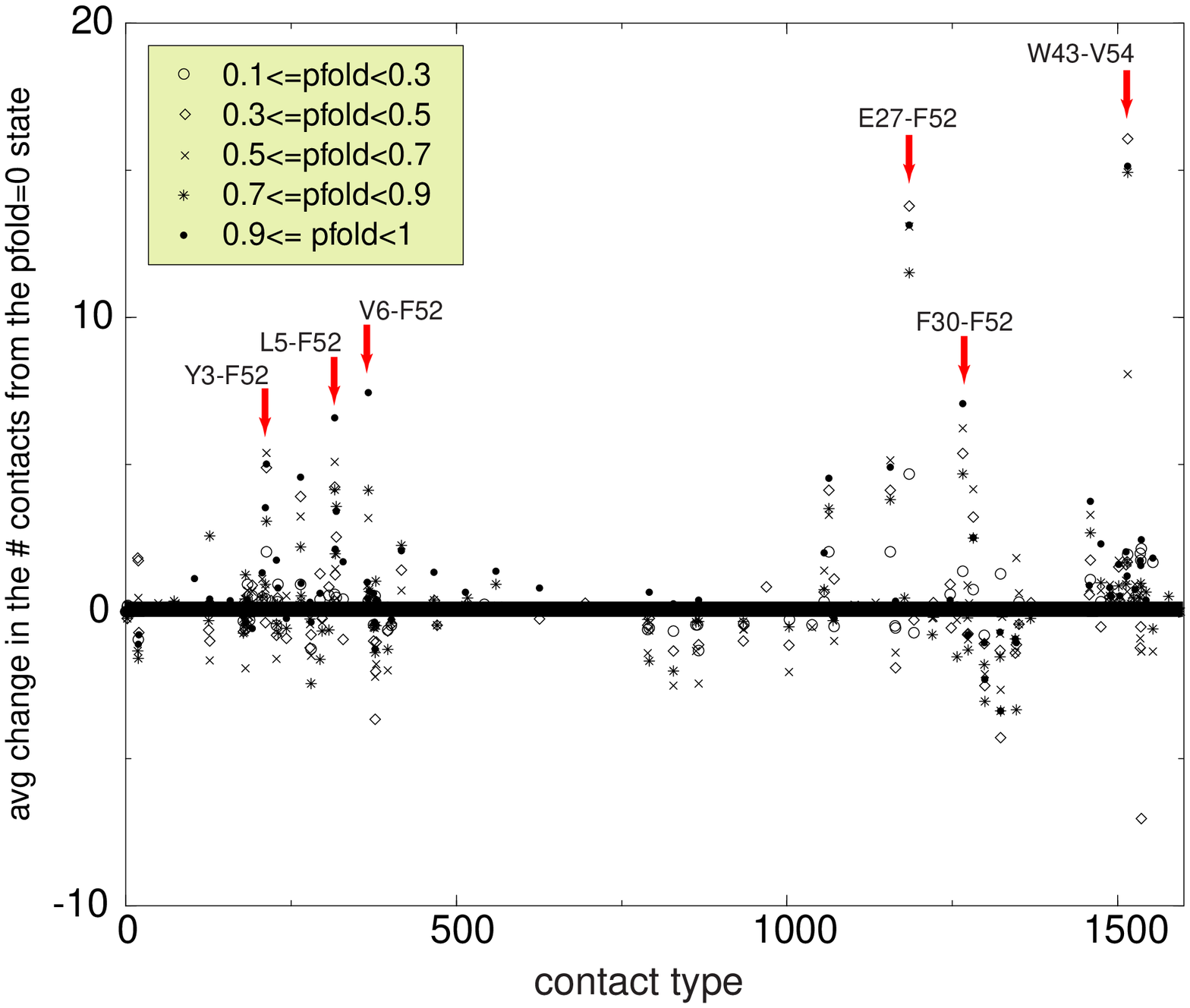,angle=90}
\end{center}
\begin{center}
\epsfig{file=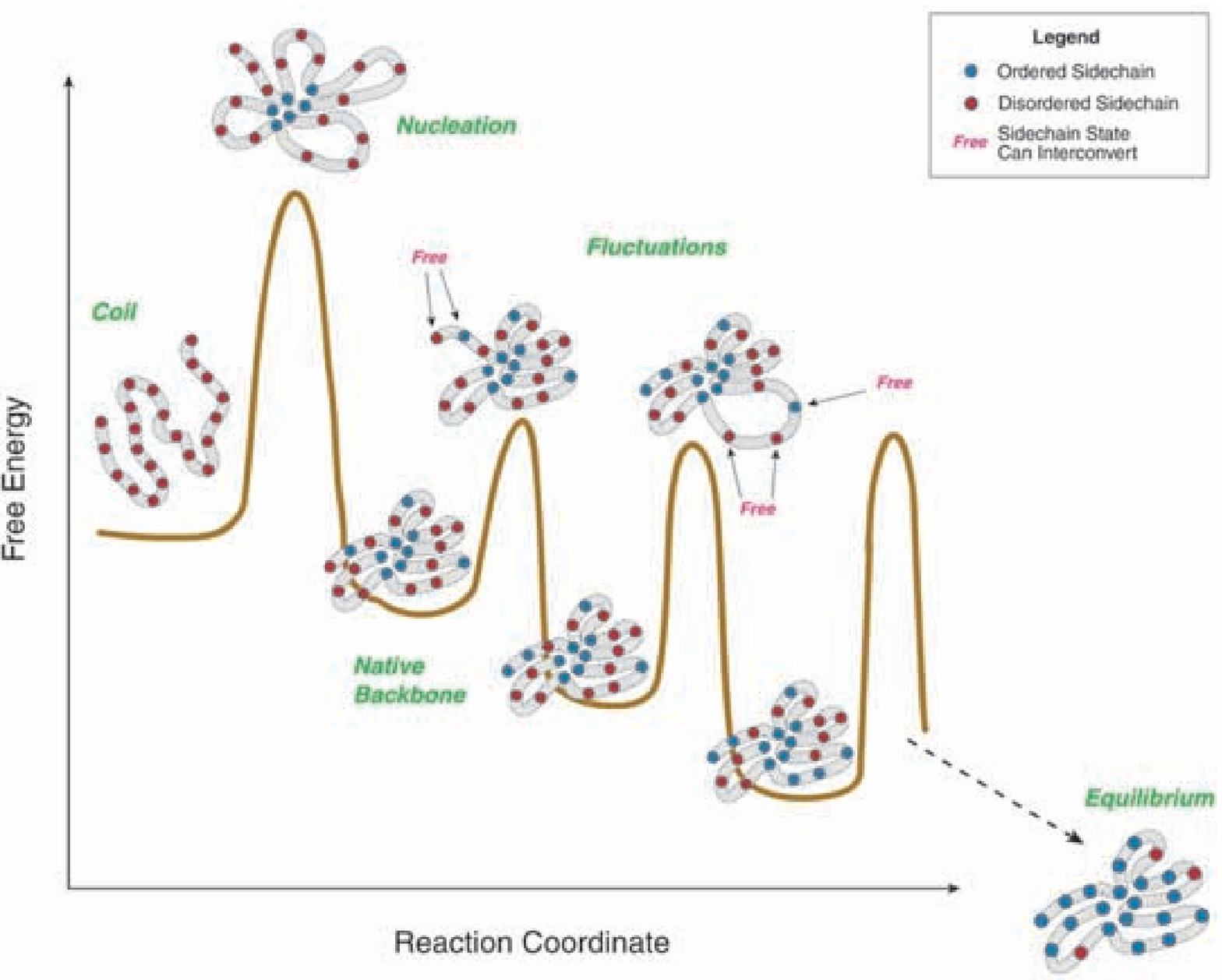,angle=90}
\end{center}
\newpage

\begin{figure}[tbp]
\caption{} \label{fig:thermo}
\end{figure}
\begin{figure}[tbp]
\caption{} \label{fig:kinetics}
\end{figure}
\begin{figure}[tbp]
\caption{} \label{fig:48mer_kin}
\end{figure}
\begin{figure}[tbp]
\caption{} \label{fig:48mer_traces}
\end{figure}
\begin{figure}[tbp]
\caption{} \label{fig:rate_hists}
\end{figure}
\begin{figure}[tbp]
\caption{} \label{fig:48mer_struct}
\end{figure}
\begin{figure}[tbp]
\caption{} \label{fig:energy_relax}
\end{figure}
\begin{figure}[tbp]
\caption{} \label{fig:igd_traces}
\end{figure}
\begin{figure}[tbp]
\caption{} \label{fig:fast_residues}
\end{figure}
\begin{figure}[tbp]
\caption{} \label{fig:three_state}
\end{figure}
\begin{figure}[tbp]
\caption{} \label{fig:residue_contacts}
\end{figure}
\begin{figure}[tbp]
\caption{} \label{fig:residue_contacts_detail}
\end{figure}
\begin{figure}[tbp]
\caption{} \label{fig:cartoon}
\end{figure}
\end{document}